\documentclass[twocolumn,twoside,slac_two]{revtex4}
\usepackage{graphicx}
\usepackage{fancyhdr}
\usepackage{color}
\begin{document}

\title{Accelerator Preparations for Muon Physics Experiments at Fermilab}\thanks{Work supported by the U.S. Department of Energy under contract No. DE-AC02-07CH11359.}

%

\author{M.J. Syphers}
\affiliation{Fermilab, Batavia, IL 60510, USA}

\begin{abstract}
The use of existing Fermilab facilities to provide beams for two muon experiments --- the Muon to Electron Conversion Experiment (Mu2e) and the New g-2 Experiment --- is under consideration.  Plans are being pursued to perform these experiments following the completion of the Tevatron Collider Run II, utilizing the beam lines and storage rings used today for antiproton accumulation without considerable reconfiguration.  Operating scenarios being investigated and anticipated accelerator improvements or reconfigurations will be presented.
\end{abstract}

\maketitle


\section{Preface}

Fermilab's neutrino program has included upgrades to the Main Injector (MI) which will allow it to run with a 1.333~sec cycle time, with twelve batches of beam from the Booster synchrotron being accumulated into the Recycler storage ring before being single-turn injected into the MI at the beginning of the MI cycle.  Recent upgrades have increased the maximum average Booster repetition rate from roughly 2.5~Hz to 9~Hz in order to meet the demands of the neutrino program.  A further upgrade to the Booster RF system to be performed over the next several years will allow the Booster to run at its maximum rate of 15~Hz.  At 15~Hz, there remain eight Booster cycles during each MI period that could in principle be used for an 8~GeV (kinetic energy) beam experimental program, with  $\sim4\times 10^{12}$ protons (4~Tp) per cycle.

One experiment, the Muon to Electron Conversion Experiment (Mu2e), has been given relatively high priority by HEPAP and Stage-I Approval from the Fermilab PAC.  The experiment will take several years to construct.  Meanwhile, during the time period between when the Tevatron Run II program is concluded and Mu2e begins, much of the same facility components can be used to furnish beam to another proposed experiment, the Muon g-2 Experiment (g-2) which could be relocated from Brookhaven National Laboratory.   In what follows we look at each experiment's beam requirements and discuss how the Fermilab complex could readily meet those needs.  The scenario proposed closely follows many of the concepts outlined in various talks and reports that have been around since 2006.\cite{b:McGinnis}  As presently understood, the g-2 experiment would be most likely to come on line first during the construction phase of the Mu2e experiment.

For either experiment beam will be transported directly from the Booster to the Recycler ring, an operation also required (and funded) for the neutrino program.  Mu2e and g-2 would also require extraction from the Recycler into the existing P1 beam line for transport toward the existing antiproton storage rings.  As the Debuncher, Accumulator, and Recycler rings are all involved in current scenarios for these experiments, stochastic cooling (and, in the Recycler, electron cooling) equipment used for antiproton production will be removed to generate less aperture restrictions for the high intensity operations of the future 8~GeV experimental program.

Particle losses in the Booster currently limit the beam delivered by this synchrotron to about $1.6\times 10^{17}$ protons/hour.  Comparatively, 15~Hz operation at 4~Tp per pulse would produce roughly $2.2\times 10^{17}$ protons per hour.  It is expected that the new magnetic corrector system \cite{b:PrebysPAC07}, the installation of which was recently completed, will eventually allow for this increased intensity under 15~Hz operation.  Measures will need be taken to improve the environmental impact of the new uses of the antiproton source storage rings under these new high intensity conditions.

\section{The Mu2e  Experiment\label{sec:mu2e}}

The Muon to Electron Conversion Experiment (Mu2e)\cite{b:mu2eProposal} requests a total delivery of $4\times 10^{20}$ protons on target (POT) per year for two years of running.  Muons are to be produced and brought onto an aluminum stopping target in narrow ($<$200~ns) time bursts, separated by intervals of about 1.5~$\mu$s, somewhat larger than the lifetime of muonic aluminum.  Muon to electron conversion data would be taken between bursts, after waiting a sufficient time ($\sim$700~ns) for the prompt background to subside.  A suppression (extinction) of the primary proton beam between bursts by a factor of $10^{-9}$ relative to the burst  itself is necessary to control the prompt background.

\subsection{Meeting Experimental Requirements}

\begin{figure}
  \includegraphics[width=0.39\textwidth]{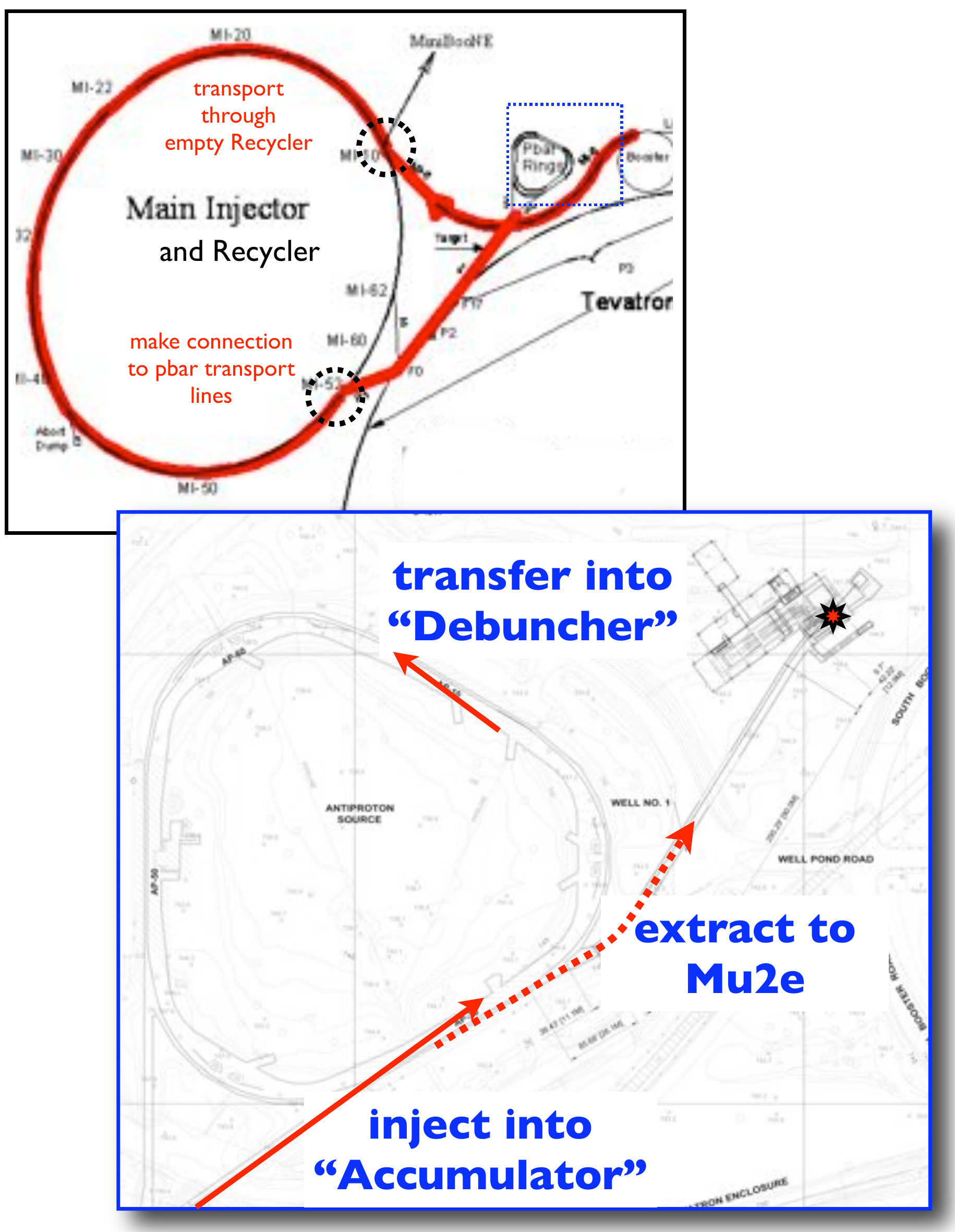}
  \caption{Beam transport scheme for Mu2e operation.}\label{f:mu2e:schem}
\end{figure}

The proton delivery method proposed by the experiment is to send Booster beam through the Recycler and directly inject into the Accumulator, where several Booster batches would be momentum stacked.  Thus, in this scenario, the Recycler is used as a simple beam transport, and the Accumulator/Debuncher rings are used to generate the desired beam properties.  Since this is carried out with 8~GeV kinetic energy proton beams, no new beam lines are required, and all magnetic elements operate at their present-day field strengths.  A schematic of the beam line system is presented in Figure~\ref{f:mu2e:schem}.

\begin{figure}
  \includegraphics[width=0.45\textwidth]{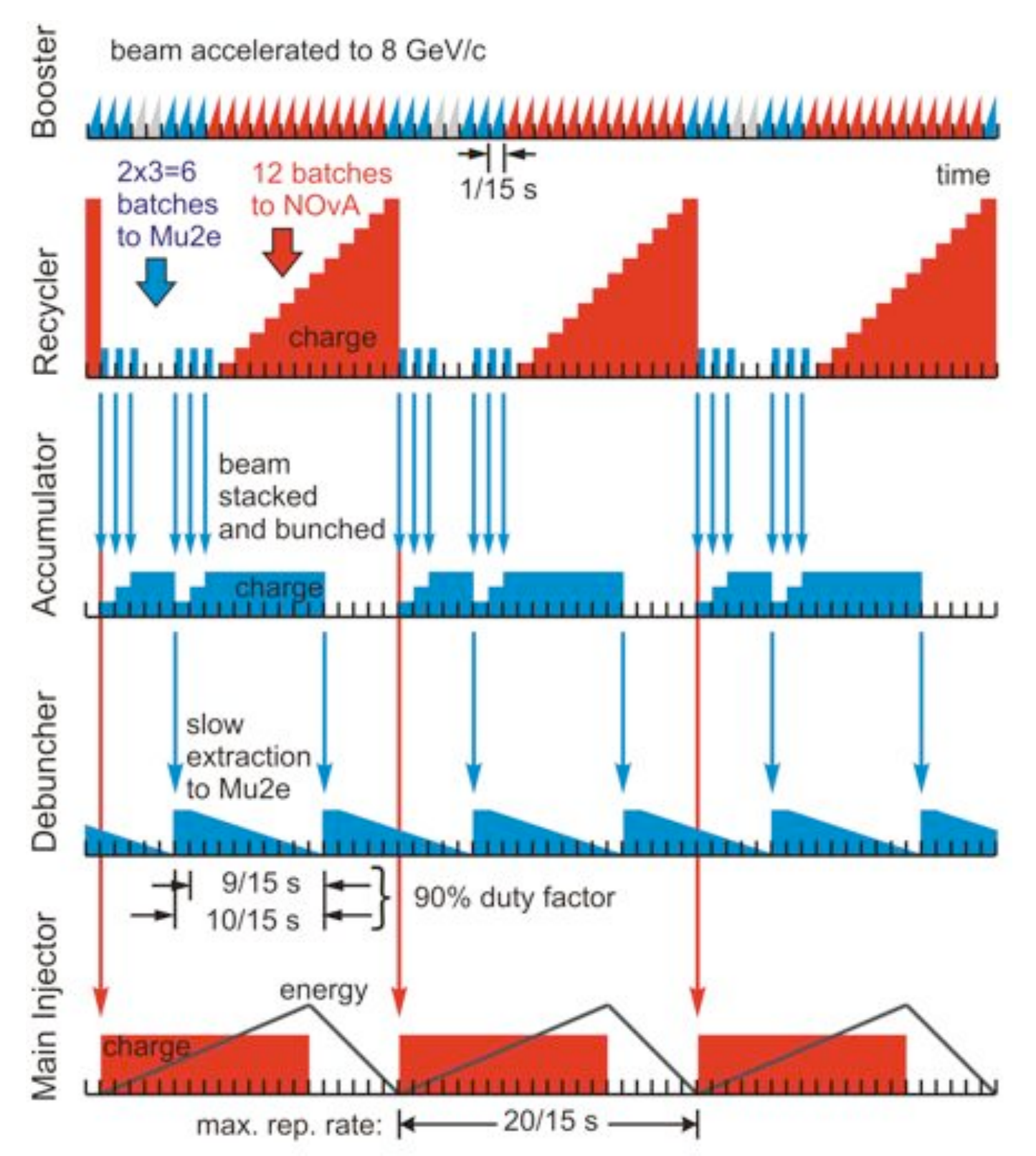}
  \caption{Proposed timing scheme for Mu2e.}\label{f:mu2e:ramps}
\end{figure}

Six of the eight free Booster cycles are used to feed 4~Tp per pulse to the Mu2e experiment, three batches at a time.  Figure~\ref{f:mu2e:ramps} shows the proposed time line of events during MI operation.
Three consecutive batches are momentum stacked into the Accumulator ring and then coalesced into a single bunch using an $h=1$ RF system.  This beam is then transferred into the Debuncher ring where a bunch rotation is performed and a single short bunch, of $\sim$40~nsec extent (rms), is captured into an $h=4$ RF system.  The total process to this point would occur within five Booster cycles.  The beam then would be resonantly extracted from the Debuncher over the next 9 Booster cycles.  This single bunch would produce a train of 40~nsec (rms) bursts being emitted from the Debuncher at 1.7~$\mu$sec intervals (the revolution period of the Debuncher ring) producing a structure well suited to the Mu2e experiment.  Beam would be transported through an 8~GeV beam line to the experiment, presumably located to the northwest of the Debuncher/Accumulator tunnel.  During this extraction from the Debuncher, the Accumulator can be re-filled with three more Booster batches to await transfer to the Debuncher.  As can be seen in Figure~\ref{f:mu2e:ramps}, a total of six batches per Main Injector cycle time of 1.33~sec can be slow spilled to the experiment with a duty factor of 90\%.  If each batch contains 4~Tp, then the Debuncher will start with 12~Tp and if spilled over 9/15~sec at 1.7~$\mu$sec per burst will yield $34\times 10^6$ protons per burst onto the target, with an average spill rate of 18~Tp/sec and a total of $1.8\times 10^{20}$ protons on target within a ``Snowmass year'' ($10^7$~s).

\subsection{Beam Preparation and Delivery}
For Mu2e, an 8~GeV proton beam must be injected from the MI-8 transport line into the Recycler, and extracted from the Recycler into the P1 transport line.  The injection line is part of the NO$\nu$A project.  The Mu2e experimental scenario described above only requires beam to circulate part-way around the Recycler.  Thus, either (a) an extraction kicker similar to that used for injection can be arranged for extraction as well, or (b) a pulsed dipole magnet can be turned on during the Booster cycles from which beam passes through the Recycler.

Once out of the Recycler and into the P1 line, the beam is transported to the Accumulator ring in the same manner as is done presently for so-called ``reverse proton'' operation (used during tune up of the antiproton source).  Hardware to transfer beam between the Accumulator and the Debuncher exists and is used routinely, though upgrades for higher repetition rates may be required.

\subsubsection{Mu2e Beam Line}

\begin{figure}
\includegraphics[width=0.48\textwidth]{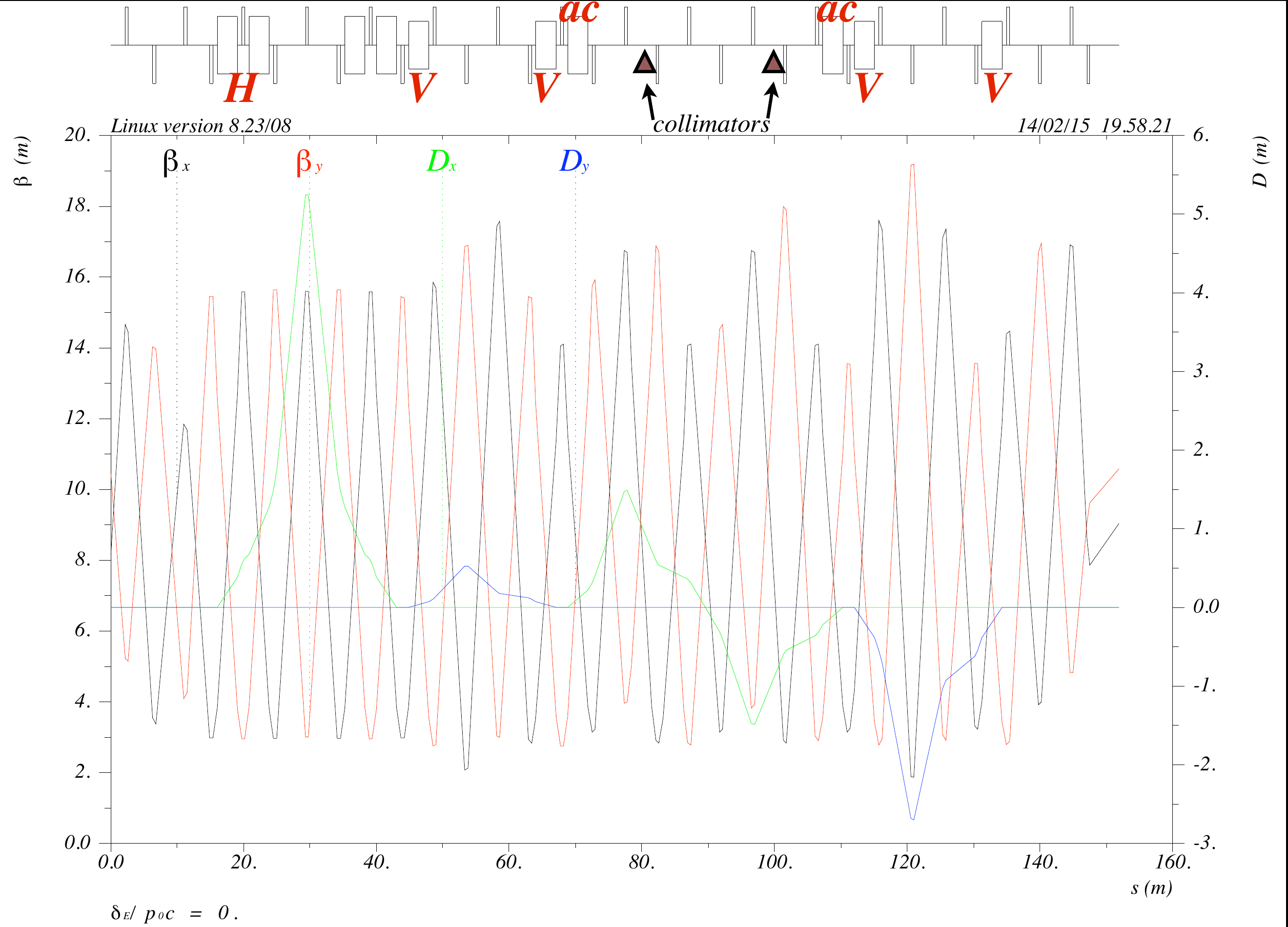}
\caption{Schematic of beam line transporting 8.9 GeV/c beam from the Debuncher to the Mu2e production target.  ``H'' and ``V'' in this preliminary design denote the locations of horizontal and vertical bending elements, and ``ac'' indicates possible locations of the AC dipole magnets used in the extinction process, with collimators located in-between.}\label{f:mu2e:line}
\end{figure}

Design effort for the extraction line leading toward the experiment has been started.  The length of the extraction line, using the layout depicted in Figure~\ref{f:mu2e:schem},  will be approximately 150-200~m and its cost and complexity roughly can be scaled from the many other 8~GeV beam lines built at Fermilab over the past decades.  While a final design for the beam line elements is not in place, it will be conceptually similar to other 8~GeV transport systems, for example the miniBooNE beam line at Fermilab.  One exceptional feature of the line is the extinction insert,  discussed separately below.  Otherwise the line will contain on the scale of 20-30 quadrupoles, a few minor bend centers, and standard cooling, powering, and instrumentation requirements.  A schematic of an early beam line design showing possible betatron optical functions is provided in Figure~\ref{f:mu2e:line}.\cite{b:CJbeam}

\subsubsection{RF Requirements}
\begin{figure}
\includegraphics[width=0.45\textwidth]{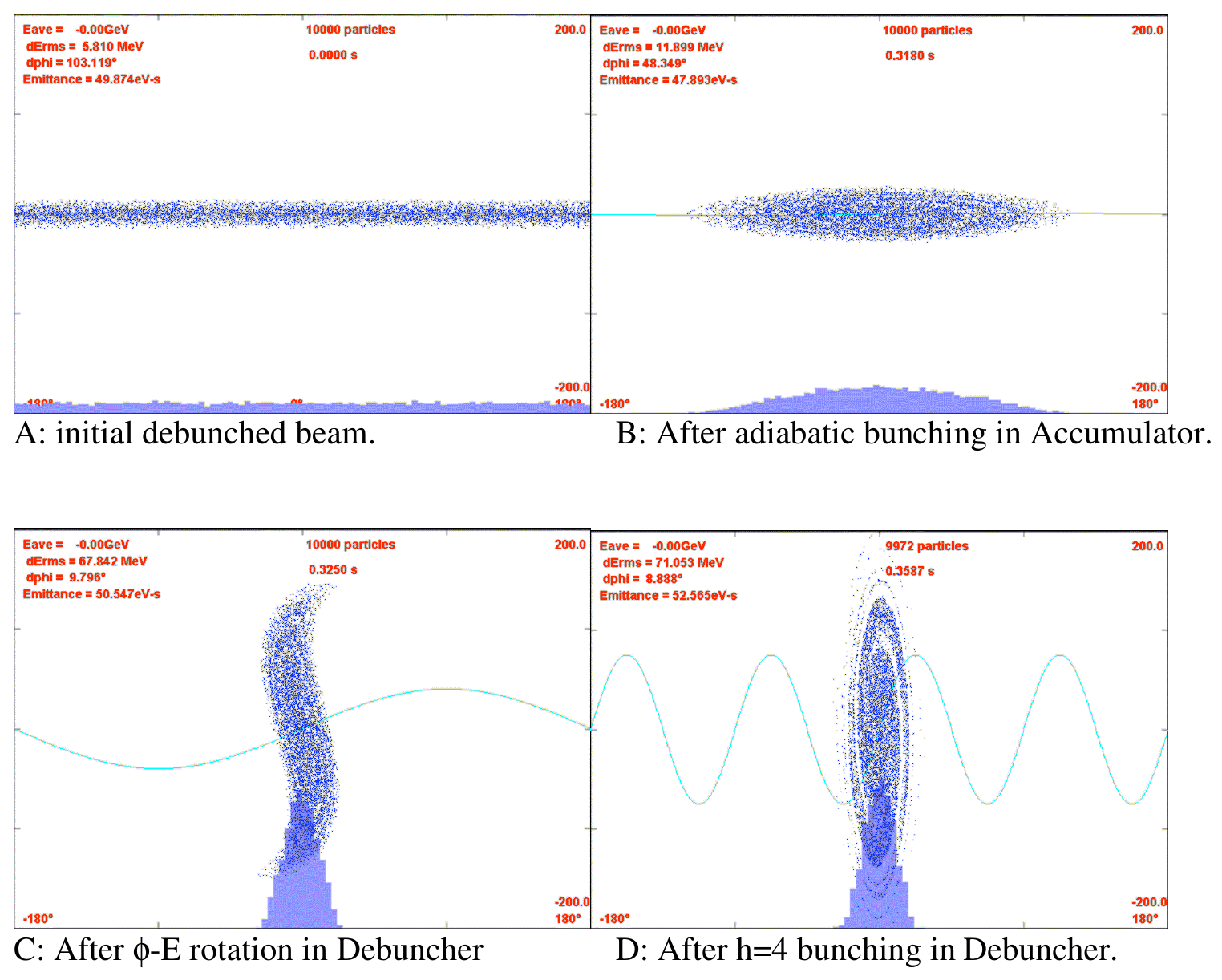}
\caption{Particle distributions in phase (horizontal, $\pm 180^\circ$ or $\sim\pm$0.85~$\mu$sec) and energy (vertical, $\pm$200~MeV) phase space, with histograms shown along the bottom edge, for stages of Mu2e bunch preparation.  Curves indicate the RF wave forms used. }\label{f:mu2e:BunchRotation}
\end{figure}

As noted earlier, the major beam preparation for the Mu2e experiment is performed in the Accumulator and Debuncher rings.  
The Accumulator with its large aperture and momentum stacking system  is well suited for accumulating pulses of protons from the Booster ({\em via} the Recycler) and stacked three at a time.    Protons enter the Accumulator onto an ``outer'' orbit, are captured with 53~MHz RF and decelerated toward the ``core'' orbit where they merge with already circulating particles.  Should the present system require more total voltage to enable three consecutive batches from the Booster to be accumulated, the  Debuncher's 53~MHz system, not needed in the new scenario, can be relocated to the Accumulator.   Once three Booster batches have been accumulated in this way, the present scheme (\cite{b:NeufferScheme}) uses an $h=1$ RF system that is turned on adiabatically to 4~kV, capturing the beam into a single bunch.  This allows enough time for an extraction kicker to fire sending the beam to the Debuncher ring.  Once in the Debuncher, a similar $h=1$ system running at 40~kV will cause the bunch to rotate in phase space, generating larger momentum spread but shorter bunch length.  After $\sim$7~msec the bunch rotates $90^\circ$ at which time it is captured by an $h=4$ RF system running at 250~kV.  This system keeps the beam bunched with an rms length of 38~nsec and energy spread of $\pm$200~MeV.  Figure~\ref{f:mu2e:BunchRotation} displays the evolution of the longitudinal phase space through the process.

The Accumulator and Debuncher rings at present contain $h=1$ and $h=4$ RF systems, but are run at much lower voltages ($<2$~kV).  Thus, upgrades to these systems will be in order, including additional cavity hardware and high level RF amplifiers.

\subsubsection{Slow Extraction}

Resonant extraction is a technique for slowly and relatively evenly removing particles from a synchrotron, and has a long history at Fermilab.  The original Main Ring, the Tevatron, and the Main Injector have all used half-integer resonant extraction for producing slow spill particle beams for targeting.  In these cases, the non-integral part of the betatron tune resides near one half, and a fast quadrupole magnet system with feedback circuitry is used to carefully ease the tune toward the half-integer.  Due to nonlinear magnetic fields inherent in any real magnet system, which can be further enhanced by the introduction of tunable octupole magnets, particles with larger betatron oscillation amplitudes will have tunes that go on-resonance first, increasing their amplitudes even further, and these particles can be directed into an extraction channel leaving the synchrotron.  As the tune slowly approaches 0.5, the higher amplitude particles are ``peeled off'' from the distribution, generating a smooth stream of particles leaving the ring.  

The Debuncher, with its three-fold symmetry and a design tune near a third of an integer, makes the use of third-integer extraction a possibly attractive option for the Mu2e application.  Here, sextupole magnets are used to enhance the resonance at a tune of 1/3 generating a dynamic aperture (or stable phase space area) that is proportional to the difference of the tune from 1/3.  As the tune adiabatically approaches 1/3, particles that suddently find themselves outside the dynamic aperture stream away from unstable fixed points in a well defined pattern and, as in the half-integer case, will eventually wander to the other side of a septum to be directed out of the synchrotron.

The exact system to be chosen requires further study.  One of the major benefits of half-integer extraction is the fact that the entire phase space can be made unstable when the tune gets close enough to 0.5 (when the beam enters the half-integer stop-band gap).  This allows for the complete removal of the particles from the synchrotron to the experiment, and is one of the primary reasons half-integer extraction was chosen for the three Fermilab synchrotrons mentioned above.  The third-integer system will have particles remaining in the ring which will need to be aborted at the end of the slow spill.  Also, when the particle beam has a large momentum spread, which will be true for either case with the Debuncher application ($\pm$200~MeV/ 8.9~GeV = $\pm$2\%), the chromaticity will need to be very finely controlled in coordination with other extraction parameters.  A major concern is the tune spread due to space charge and its effect on the extraction process.  All of these considerations are being actively investigated, though the list of necessary components is relatively well understood.

\subsection{Extinction}

As $34\times 10^6$ protons on average should reach the production target every micropulse (every 1.7~$\mu$s) during the appropriate time window, an extinction at the level of $10^{-9}$ permits no more than 1 proton to reach the target during the measurement time window over 30 micropulses.  With this stringent of a requirement, several measures must be taken to ensure the appropriate level of extinction.

\subsubsection{Internal Extinction}

Measures will be taken within the rings during bunch formation to abate particles from being outside the $\pm 100$~ns time window of the production micropulses.  For example, tight rise and fall time requirements for the kicker magnets used in transferring the bunch from the Accumulator to the Debuncher will help.  In the Debuncher ring itself, a gap-cleaning kicker system may be employed.  Also, as the narrow bunch length will necessarily produce a bunch with large momentum spread, a collimator system at a large dispersion point in the ring can also be used to scrape away particles before they migrate between stable fixed points of the $h=4$ RF system.

\subsubsection{External Extinction}

\begin{figure}
\includegraphics[width=0.48\textwidth]{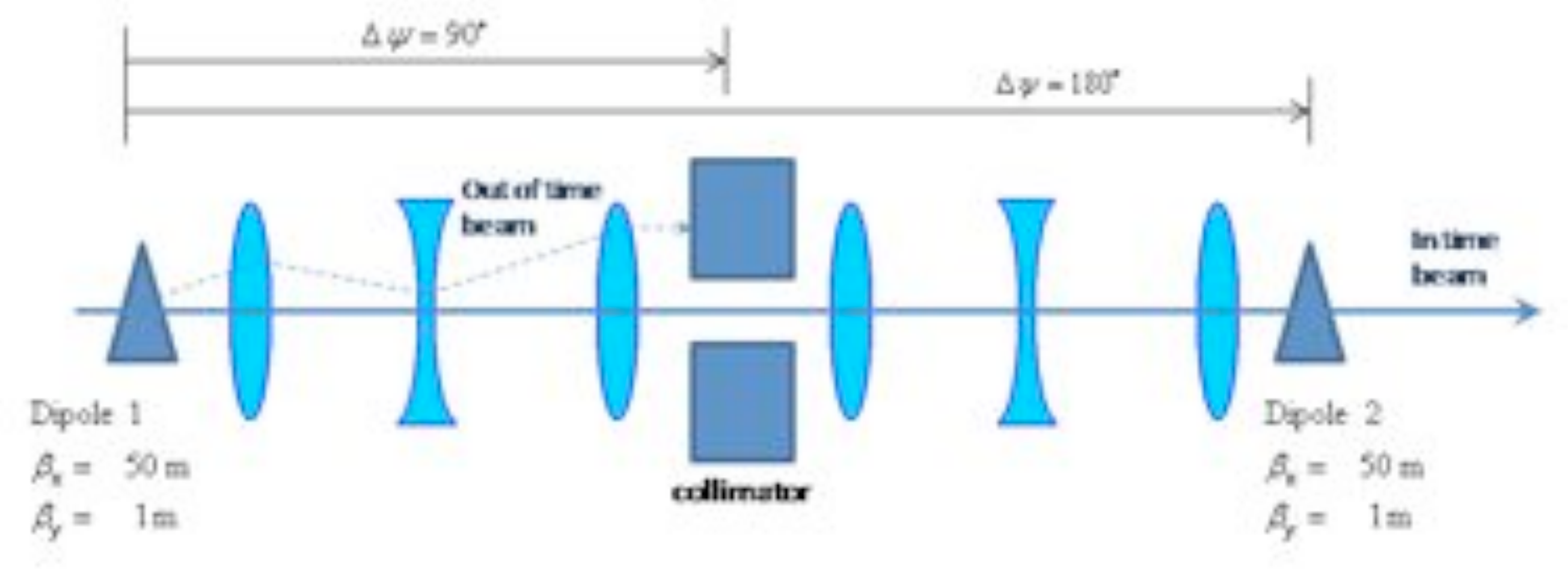}
\caption{Schematic of a possible extinction insert.  Two AC dipoles steer the ideal trajectory into collimators at an oscillation frequency of 300~kHz.}\label{f:mu2e:extinction}
\end{figure}

\begin{figure}
\includegraphics[width=0.48\textwidth]{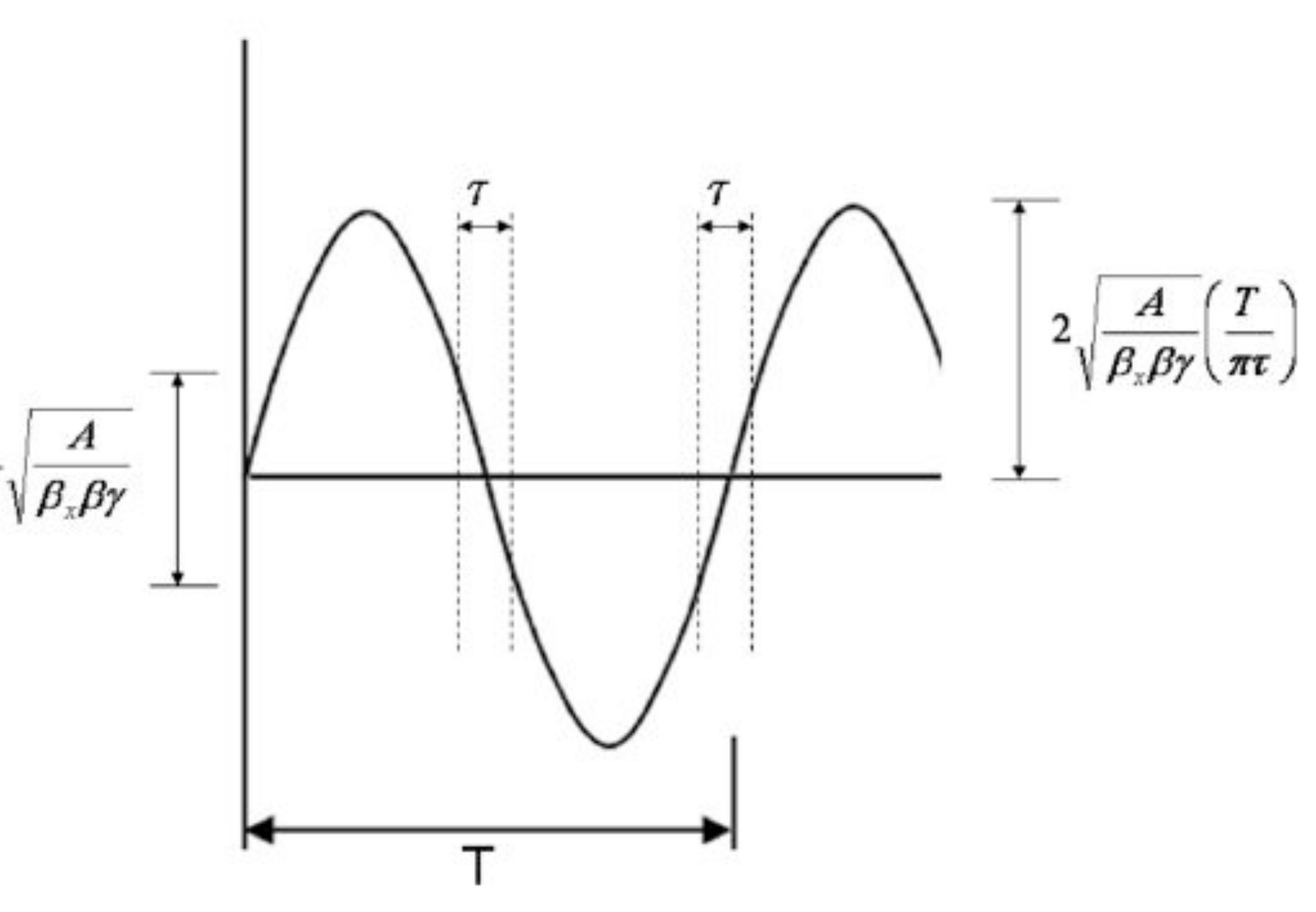}
\caption{Relationships between dipole oscillation period $T$, in-time window $\tau$, transverse admittance $A$, and betatron amplitude function $\beta_x$ at the location of the dipoles, for a two-dipole solution.  See  \cite{b:PrebysAC}.}\label{f:mu2e:ACwave}
\end{figure}

In addition to the above, the Mu2e beam line will include an ``extinction insert''  at its downstream end.  This portion of the transport system, the ``last resort'' for the extinction process, will utilize a rapid cycling dipole magnet (AC dipole) or a set of dipoles on either side of a focusing channel to be used to steer beam into  collimators.  Were a single dipole magnet used, the frequency would need to be $\sim$600~kHz (the micropulse frequency).  For a pair of bend centers, the dipole magnets would cycle at half the micropulse frequency ($\sim$300~kHz) and kick the unwanted beam well into the collimator iron. \cite{b:PrebysAC} The conceptual layout is shown schematically in Figure~\ref{f:mu2e:extinction}.  
Figure~\ref{f:mu2e:ACwave} indicates the extent of the in-time window relative to the micropulse period, $T$.  By proper choice of frequency (or set of frequencies) and amplitude, the window for particles that arrive at the production target can be adjusted.  Various hardware options for this magnet system are being explored. \cite{b:PrebysPAC09}  

\subsection{Alternate Operating Scenarios}

An inherent issue with the Baseline operating scenario described early in this document is the final bunch current in the Debuncher during the slow spill.  With a total of 3 Booster cycles accumulated into a single 12~Tp bunch with rms bunch length of $\sim$38~ns, the space charge tune shift using expected transverse emittances is roughly $\Delta\nu\sim 0.1$.  This large spread in betatron tunes, which will vary as the beam is slow spilled, will make difficulties in the resonant extraction process more pronounced.  Additionally, the highest intensity stored in the Accumulator to date is under 3~Tp (antiprotons).  An intensity higher by more than a factor of four, while not inherently impossible, will be challenging.

Alternative scenarios are being investigate which attempt to lower the bunch charge and total intensity in the rings while still providing a high duty factor to the experiment and a similar average rate of protons to target.  One obvious step is to form four bunches in the Accumulator rather than one -- reducing the space charge per bunch by a factor of four -- and transferring one bunch at a time into the Debuncher for extraction.  The spill time would be reduced from 600~ms to about 150~ms, occurring 8 times during a Main Injector cycle.  The single-turn transfers from the Accumulator to the Debuncher will also help with the inter-bunch extinction.  Several other scenarios similar to this are being investigated which have the potential to reduce the space charge even further, and may also add operational flexibility to the program.

\section{The Muon g-2 Experiment\label{sec:gminus2}}


The New g-2 Experiment\cite{b:g2Proposal} requires 3.09~GeV/c muons%
\footnote{This momentum is set not just by the already existing ring, but is a special momentum for the muon which makes the spin precession independent of electric fields.  (See \cite{b:g2PRL}.)}
injected into an existing muon storage ring that would be relocated from Brookhaven National Laboratory to Fermilab.  The muon storage ring is 7~m in radius, giving a revolution time of 147~ns.\cite{b:g2PRL}  To account for the injection kicker, the beam  pulses need to have lengths of about 100~ns or less.  These pulses should be separated on the scale of about 10~ms for the muons to decay in the ring and data to be recorded prior to the next injection.  To obtain as pure a muon beam as possible entering the storage ring, the experiment would like a decay channel off of the production target that corresponds to several pion decay lengths = 7.8~m~$\times \gamma$ = 7.8~m~$\times$ 3.09/0.14 = 7.8~m~$\times 22 = 170$~m.  Present understanding of the pion yield off of an 8 GeV target at Fermilab dictates the desire to deliver a total of  $2\times 10^{20}$ 8 GeV protons on target to obtain $21\times$ more statistics for the g-2 experiment and give a 0.1 ppm measurement of the muon anomalous magnetic moment.

\subsection{Meeting Experimental Requirements}

To meet the above requirements it is envisioned that six Booster batches every MI cycle can be sent to the experiment for an average rate of 6/20 $\times$ 4~Tp $\times$ 15/sec = 18~Tp/sec.  This yields the required total protons on target in about a single ``Snowmass year'' of running.  Each batch of 53~Mhz bunches from the Booster would be sent to the Recycler and coalesced into four bunches for delivery to the experiment.  Using existing RF systems, possibly supplemented with like-kind components, the four bunches can be formed to meet the demands of the g-2 ring.   The re-bunching process takes approximately 30~ms, and the four bunches would then be delivered to the experiment one at a time spaced by 12~ms.  Thus, the last bunch is extracted just within the 66.7~ms Booster cycle.  The remaining two Booster cycles, before and after this process, allow for pre-pulsing of fast devices prior to the change between NuMI and ``muon'' cycles.  (If this is deemed unnecessary, then eight rather than six Booster cycles could feed the experiment during each MI cycle.)  

\begin{figure}
\includegraphics[width=0.48\textwidth]{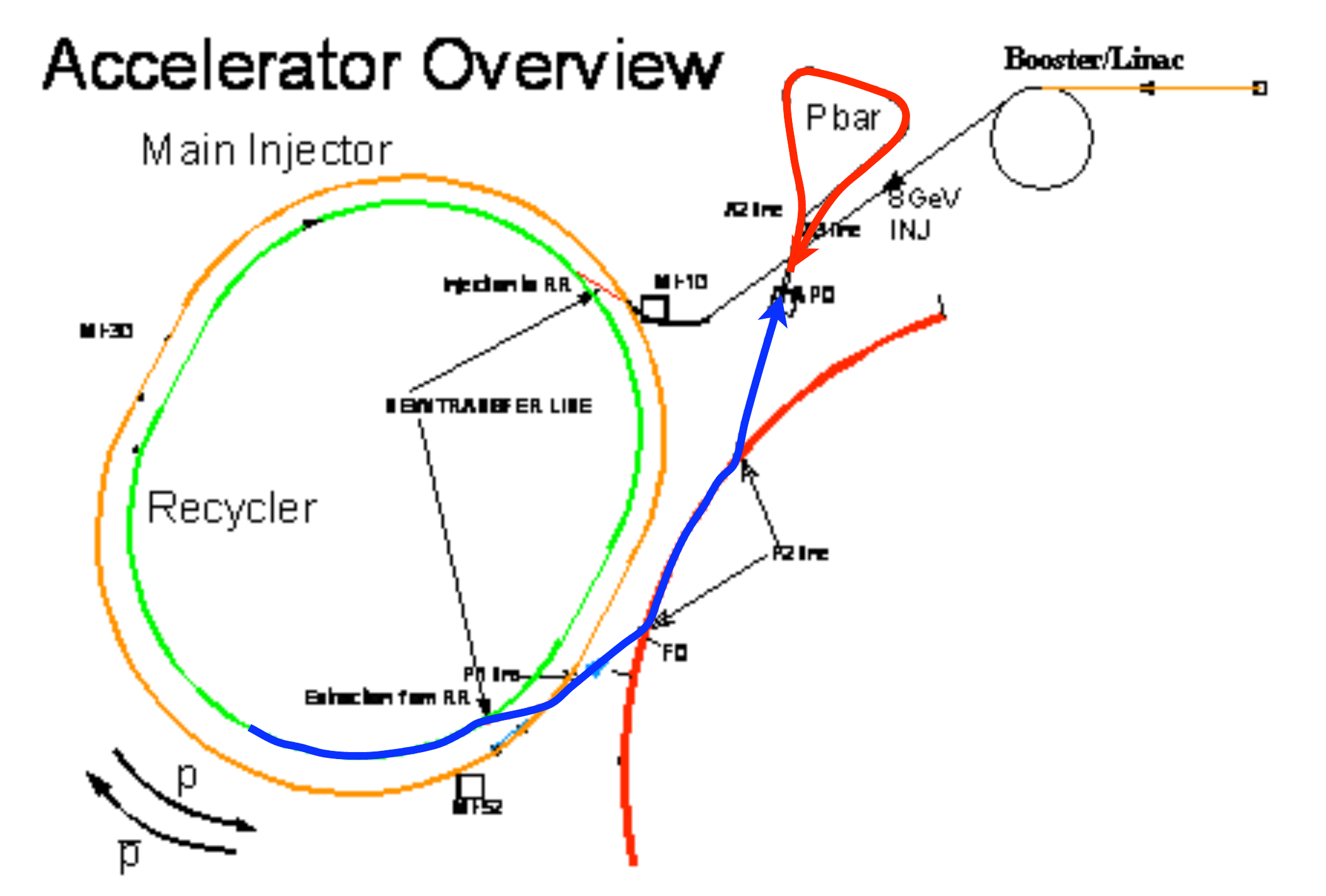}
\caption{Beam transport scheme for g-2 operation.  Beam is prepared in the Recycler, exits via the P1 line, passes through the Tevatron tunnel into the AP1 beam line, and to the AP0 target area. (Blue curve.)  Pions, decaying to muons, are transported from the target through the AP2 line, once around the ``pbar'' rings (Debuncher/Accumulator) and back toward the experimental hall near AP0 via the AP3 beam line. (Thick red curve.)}\label{f:g-2:schem}
\end{figure}

Once extracted from the Recycler a bunch is sent toward the existing, though possibly modified, antiproton target station for $\sim$3.09~GeV/c pion production.  A ``boomerang'' approach utilizing the Debuncher and Accumulator rings can be used as a delay line allowing for pion to muon decay, assuming a final location of the g-2 ring in the vicinity of the production target.  A schematic of the beam line system is presented in Figure~\ref{f:g-2:schem}.
The total length of the decay line would be $\sim$900~m.  To obtain even further purity of the muon beam, multiple revolutions in the Debuncher or Accumulator rings could be considered, perhaps as an upgrade to the program.  This upgrade would require the development of an appropriate kicker system and is not included in this first design iteration.  The 900~m decay length, however, is already a large improvement over the original decay region at BNL.

\subsection{Beam Preparation}
The major proton beam preparation will be performed in the Recycler ring.  A broadband RF system like that already installed in the Recycler would be used, except twice the voltage may be required.  The 2.5~MHz (max. $V_{rf}$ = 60~kV) and 5 MHz (max. $V_{rf}$ = 15~kV) RF systems that presently reside in the MI would be relocated to the Recycler.  Upgrades to increase their maximum voltages by roughly 10-30\% may be required.

\begin{figure}
\includegraphics[width=0.48\textwidth]{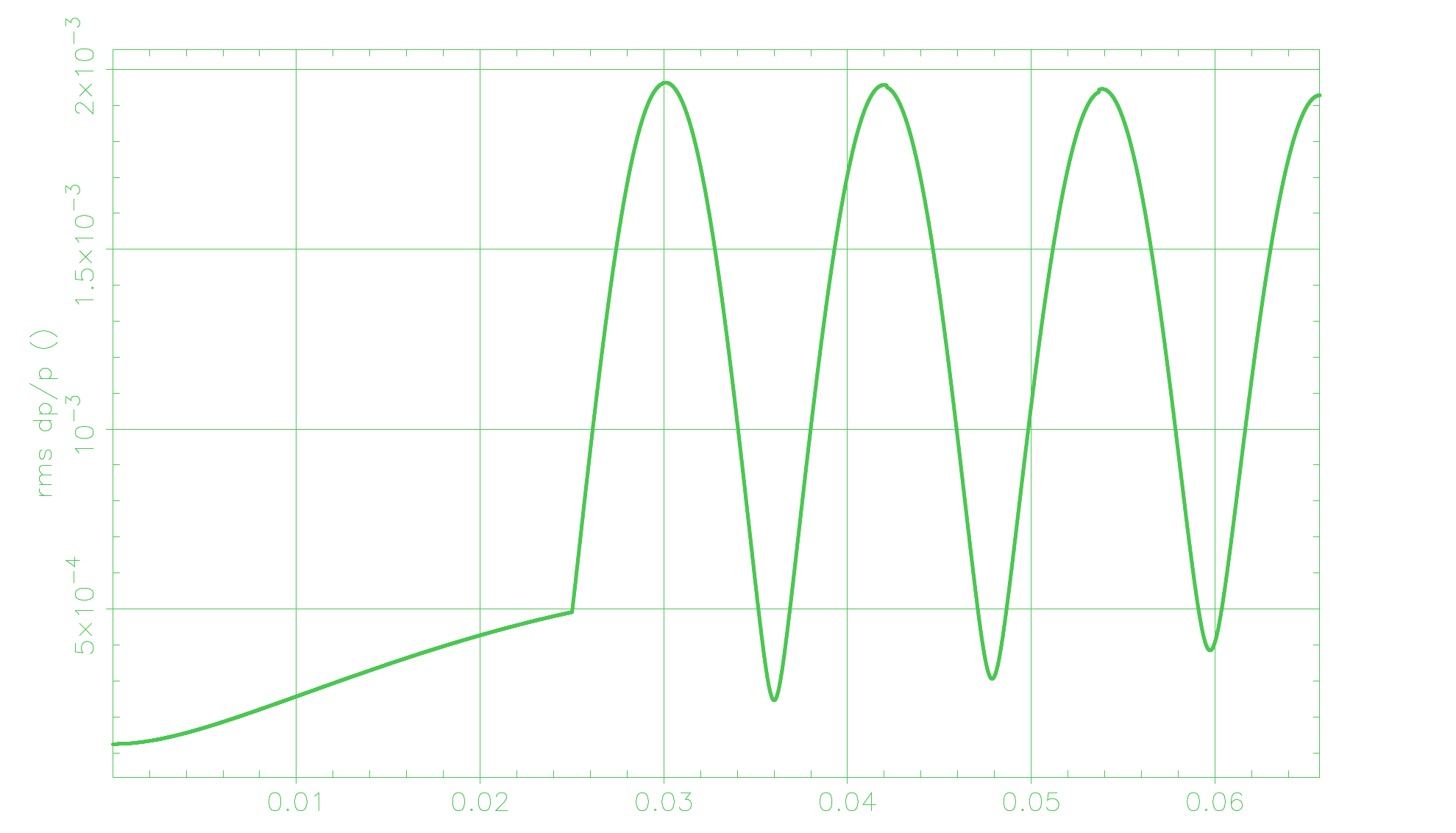}
\caption{Resulting relative momentum spread ($\Delta p/p$) {\em vs.}  time in seconds following injection into the Recycler.  After an initial phase using the broadband RF system, beam is captured into four buckets.  The beam rotates within the four buckets with period 12~msec and is extracted one-by-one as the momentum spread reaches its peak (pulse length is at its shortest).}\label{f:g-2:bunching}
\end{figure}

As described in \cite{b:BhatMac}, the bunching scheme is to use a four period sawtooth wave form across the Booster batch produced by the broadband RF system to break the batch into four segments and rotate them in phase space sufficiently that they can be captured cleanly in a linearized bucket provided by the resonant RF.  Each of the four resulting bunches is $\sim$100~ns long.  The first bunch is extracted immediately and the latter three are extracted sequentially at half periods of the synchrotron oscillation.  The beam loading of the resonant cavities will be considerable, and further details need to be considered.  It is plausible to expect that a feedforward system can be developed without serious difÞculty.   A combination of feedback with feed forward is potentially better yet, but feedforward will be required with or without feedback.   Figure~\ref{f:g-2:bunching} 
shows the resulting beam structure in the Recycler if the beam were not extracted.  The plan would be to extract one pulse at a time, every 12~msec, when the bunches are at their narrowest time extent
(4$\sigma$ widths of 38-58~nsec). The four bunches would be separated by roughly 400~nsec center-to-center.  For the sequence shown, the RF systems require voltages of 4~kV (broadband), 80~kV (2.5~MHz), and 16~kV (5.0 MHz).  A longitudinal emittance of 0.07 eV-sec per 53~MHz Booster bunch was assumed.

\subsection{Beam Delivery and Transfer}
Following the beam trajectory starting with extraction from the Booster, we see that the proton beam needs to be injected into the Recycler from the MI-8 beam line at the MI-10 region of the Main Injector tunnel.  This maneuver will be facilitated through the NO$\nu$A project, which requires the same injection procedure.   Once prepared with the RF systems as described above, the beam will need to be extracted from the Recycler and injected into the P1 beam line.  The extraction occurs at the MI-52 tunnel location, where the Main Injector ties into this same beam line.  (See Figure~\ref{f:g-2:schem}.)   The P1 beam line is used to deliver 8~GeV antiprotons from the Accumulator into the Main Injector (and on into the Recycler) in the reverse direction.  During the g-2 operation, however,  the Main Injector will contain beam destined for NuMI and so this region will need to be modified in almost exactly the same way as MI-10 to transport protons directly into the P1 line from the Recycler. 

An appropriate kicker system will also be required for this region to extract one-by-one the four proton bunches from the Recycler.  The four bunches will be separated by approximately 200~nsec,   so the kicker must rise in $\sim$180~nsec  and have a flat top of $\sim$50~nsec.  A preliminary design and cost estimate has been developed for a kicker power supply meeting the g-2 specifications.

From the entrance of the P1 line through the Tevatron injection Lambertson (which is kept off during this operation) the beam is directed through the P2 line (physically located in the Tevatron tunnel) and into the AP1 line toward the AP0 target hall.  Again, since this system is run at 8~GeV for antiproton operations, no modifications are required for beam transport in g-2 operations.  After targeting, which is discussed in the next subsection, 3.09~GeV/c pions are collected into the AP2 line which is ``retuned'' to operate at 3.09~GeV/c rather than today's 8.89~GeV/c antiproton operation.

\subsection{Target Station}

Two options are being explored at this time for meeting the targeting requirements of the experiment, and this remains as a major R\&D area for the proposal.  Using the existing AP0 target more-or-less ``as is'' is the most straightforward approach, and remains as {\em Plan A} for the experiment.  The present system is used for selecting 8.9 GeV/c antiprotons from a 120~GeV/c primary proton beam.  For g-2 one would select$\sim$ 3.09~GeV/c pions from 8.9~GeV/c primary protons by re-tuning the beam line elements located upstream and downstream of the target.  The major hurdle to overcome with this particular scenario is the much higher repetition rate required of the lithium lens and pulsed momentum selection dipole used for antiproton production.  

The lens, which pulses approximately 62~kA into the primary of a current transformer (x8 in the secondary coil), produces a 400~$\mu$s-wide pulse every 2.2 seconds at its present operating condition.  The heat load on the lens system from Ohmic heating at this current is $\sim$4~kW, while the heating from the beam -- operating at a beam power of 70-75~kW -- is $\sim$2~kW.  The capacity of the system is on the order of 10-11~kW.

Simply scaling the beam line elements from 8.9 GeV/c operation to 3.1 GeV/c operation yields 1/8 of today's power load.  Scaling from these conditions to the g-2 experiment's baseline of 18 Hz, with 25 kW beam power, the total heat load would be 20 kW, or twice the system capacity.  Reducing the power by another factor of two (or, current by another 30\%), would reduce the total heat load to just under 10~kW, within the present system's capabilities.  There is some adjustment possible in cooling water flow as well.

At this further-reduced current, it is estimated that the pion yield would be reduced from the original estimates by 27\%, assuming the lens is positioned further from the target at its longer focal length.  Although existing hardware with modified power supplies would be used, the run-time of the experiment would be extended by $\sim$33\%, all else being held constant, in this scenario.  

A secondary {\em Plan B} is also being developed for the target area for g-2 in which the pulsed lens and magnet system would be replaced by a set of radiation-hard magnetic elements that could run at DC currents.  This was the approach taken at BNL for g-2, where the average beam power on target was three times higher and in which the magnets survived this radiation environment for the life of the experiment there.   Preliminary optics and layout using BNL-style quadrupoles have been developed, and work continues in this area.  Advantage can be taken of the existing designs, drawings and actual costs of the BNL quadrupoles used for g-2, once a preferred optical layout is obtained.

Both plans are likely similar in cost, where one is trading low magnet costs but high power supply costs in Plan A for larger magnet costs but lower power supply costs in Plan B.  A major benefit of Plan B would be the potential for better reliability and less maintenance for a DC system.  But Plan A is a better understood system at this point and presents only a small hit on run time.  It could turn out to be appropriate to begin g-2 running with a Plan A target system using mostly lens equipment that exists at the end of Run II, followed by an early upgrade to Plan B for full production running.

\subsection{Pion Decay Channel}

To obtain a long decay channel for the pions off the target, the beam is transported through the AP2 beam line, into the Debuncher ring, and extracted into the AP3 line, directed back toward AP0.  (See Figure~\ref{f:g-2:schem} again.)  The Debuncher ring can be ``partially powered'' using only those magnet strings required to perform the ``boomerang''.  Either corrector magnets or DC powered trim magnets will be used in place of kickers to perform the injection/extraction.  To enhance the pion collection and muon beam transport, optics modifications will be performed on the AP2 and AP3 beam lines.

\subsubsection{Target to Debuncher}

The current AP2 lattice has a large transverse (unnormalized) acceptance which matches well to either the Li lens or the quadrupole collection of pions from the target region.  For the generation of muons with momentum 3.094 GeV/c  in pure forward decay kinematics, an initial pion momentum higher by a factor of 1.005 is required.\cite{b:g2Proposal}  However, pions with 3-4\% higher momenta decaying with muon angles of several mrad can still contribute to the magic muon flux.  Thus, in order to increase the muon flux, (i) the FODO beta function should be decreased in the decay region and (ii) a momentum bin of  ±2\% should be accepted by the lattice.

Modification (i) reduces the pion beam size in the decay FODO, so that larger decay angles still remain within the acceptance of the g-2 ring. The momentum acceptance is limited by second-order chromatic effects, where a large aperture Q1, Q2 collection system is likely to be less efficient than the Li lens, which demonstrated 2.5\% momentum acceptance as an antiproton source. One should emphasize that these FODO changes are required only for the decay region, because their importance is weighted by $e^{-z/L}$, where $L$ = 173 m is the decay length of pions at the desired momentum.  Thus, reducing the beta function in the first ~150Êm long straight section of the AP2 line is most important, followed by the second straight FODO extending to 290 m, before the beam enters the Debuncher.  From this point on, only a small fraction of muons are produced and so no changes to the Debuncher lattice are required downstream.

At present, tripling the number of quadrupole magnets in the beginning of the AP2 line is being considered. This lattice has been simulated, confirming that accepted muon flux is approximately increased 3-fold compared to the existing AP2.  More systematic studies including Plan A and B, different lattice spacings, denser lattice in the second AP2 straight section, including wider momentum bins, and others, are currently being intensively studied.

\subsubsection{Debuncher to Ring}

\begin{figure}
\includegraphics[width=0.48\textwidth]{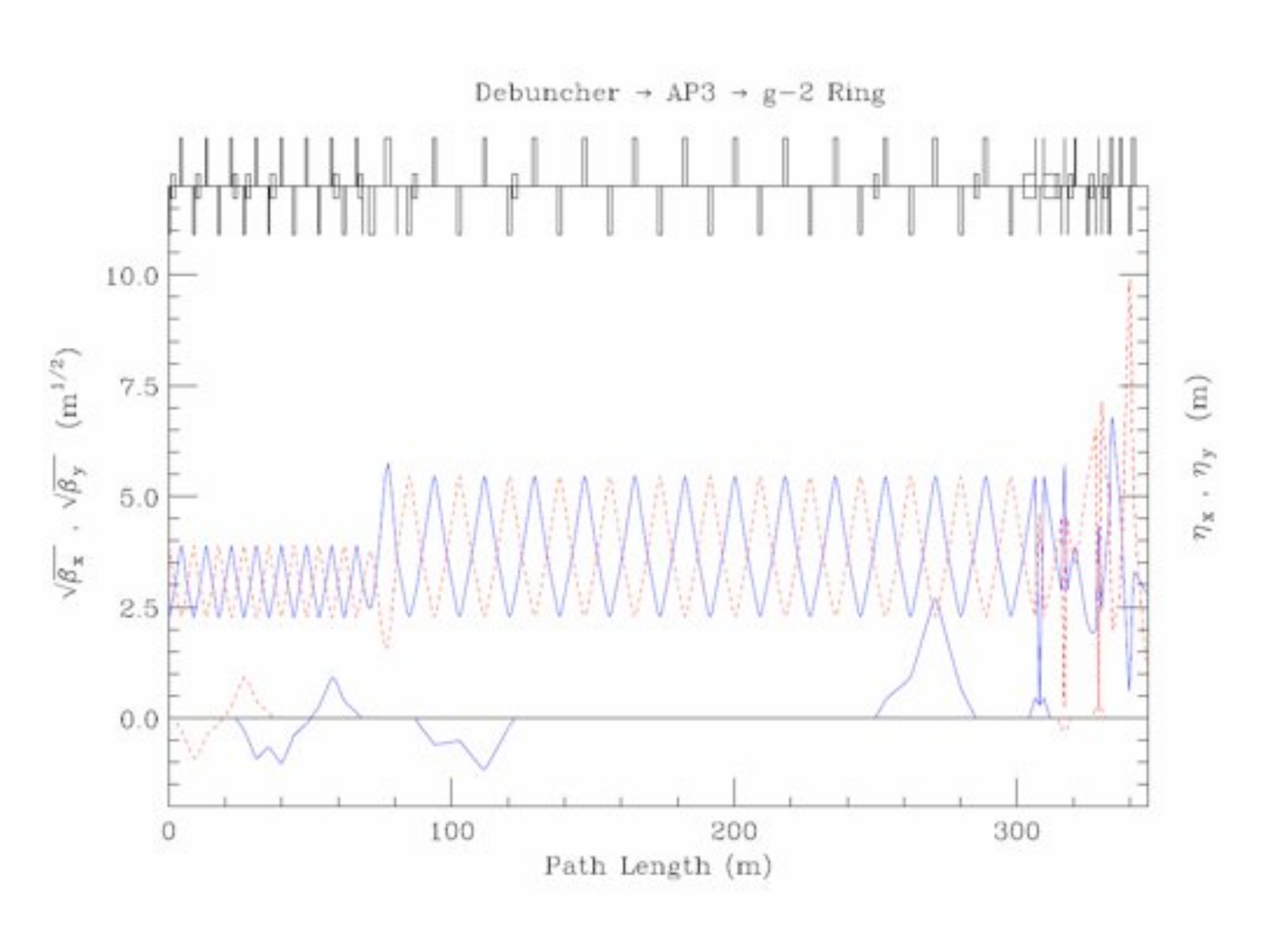}
\caption{Layout of AP3 beam line used for costing purposes.  The left axis (upper curves) indicates the relative size of the beam envelope and the right axis (lower curves) is the momentum dispersion (orbit spread per $\Delta p/p$).}\label{f:ap3line}
\end{figure}

The large aperture and strong focusing of the Debuncher is ideal for collection and transport of the pion/muon beam.  However, the AP3 line -- used today as an antiproton transport line from the Accumulator ring to the Main Injector -- has much weaker focusing and hence its admittance is much less than that of the Debuncher.  A preliminary layout of an AP3 line optics to have the same admittance as the Debuncher ring has been generated for costing purposes, as shown in Figure~\ref{f:ap3line}.\cite{b:JJline}
At the very end of the line, the beam is directed over and upward to the g-2 storage ring, and the beam is focused to a waist as it enters the inflector of the ring.  The geometry of this beam line is consistent with the geometry of the existing Debuncher ring, AP3 line and the newly designed beam line enclosure leading to the ring as described in Section~\ref{ss:bldg}.  The line is also being designed such that a bending element can be turned off and the line re-tuned for use in the Mu2e configuration (8.9 GeV/c).  The improved optics will allow for better beam transmission for Mu2e as well.

\begin{figure}
\includegraphics[width=0.48\textwidth]{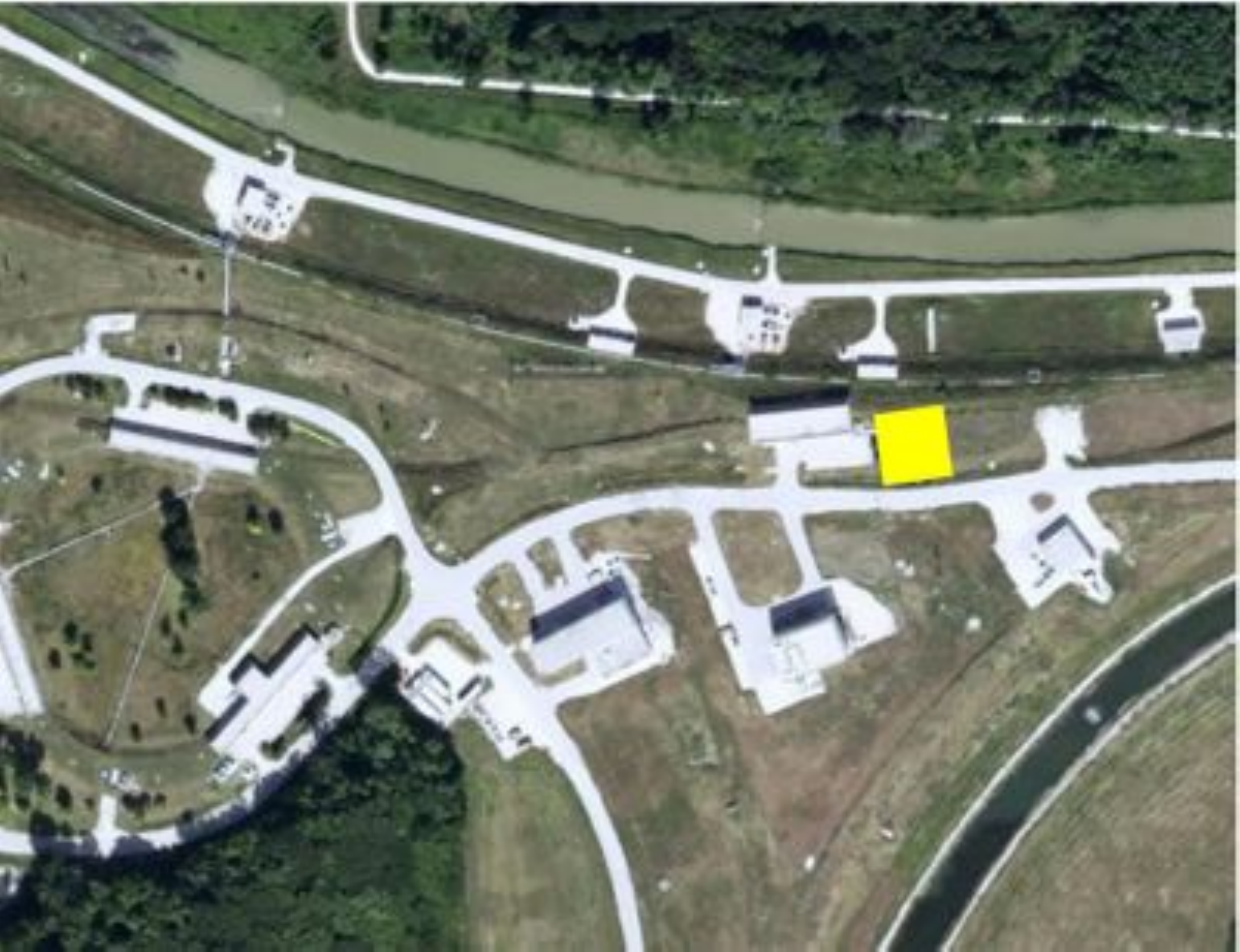}
\caption{Proposed location of the new g-2 experimental hall (yellow) at Fermilab.}\label{f:g-2:AP0goog}
\end{figure}

It is currently envisioned that the g-2 ring will be located on the surface in a new building to be constructed near the AP0 service building as indicated in Figure~\ref{f:g-2:AP0goog}.

\subsection{Experimental Facility\label{ss:bldg}}

\begin{figure}
\includegraphics[width=0.48\textwidth]{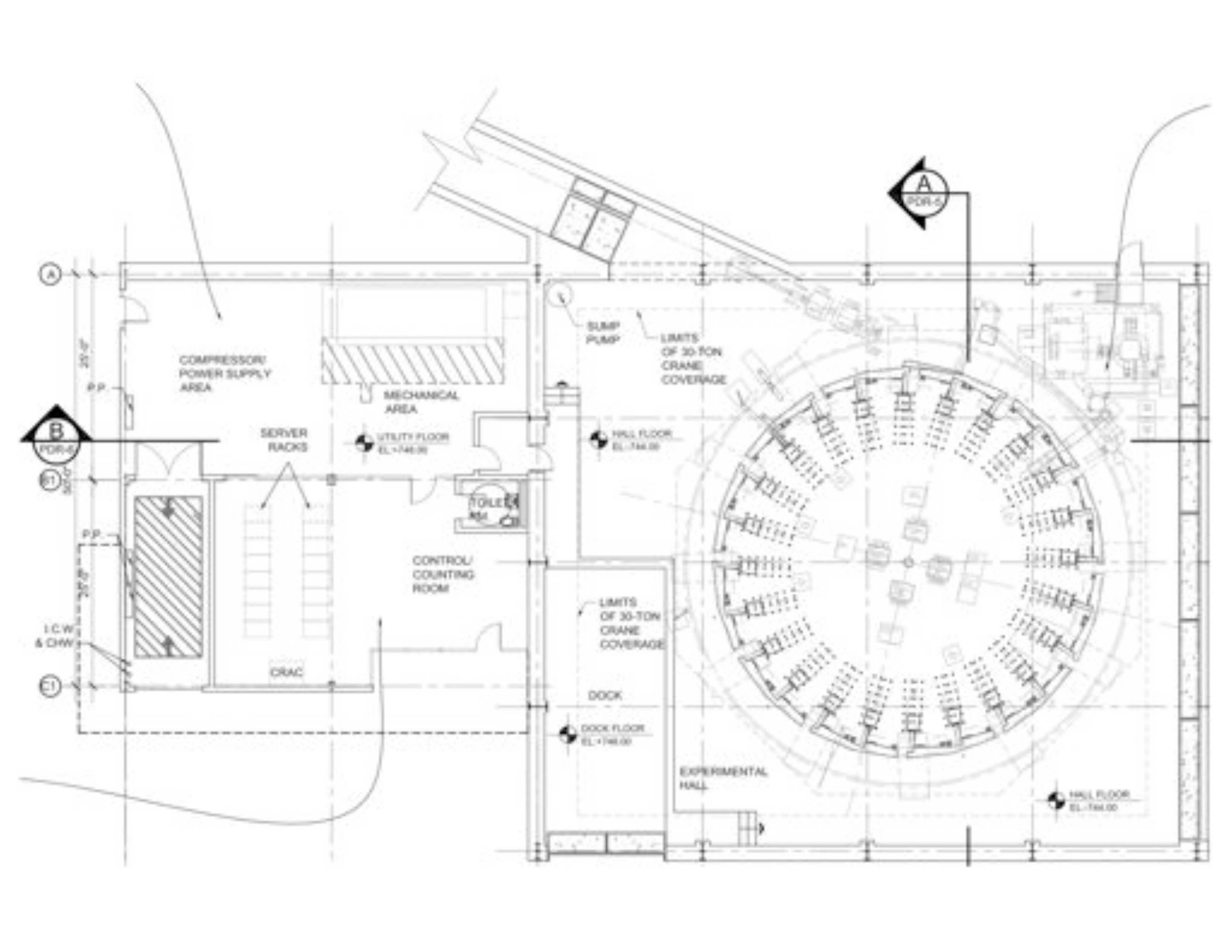}
\caption{Layout of proposed FNAL g-2 experiment hall.}\label{f:g-2:g2Hall}
\end{figure}

Considerable work has been performed at Fermilab on the design of an experimental building and beam line connection for g-2.\cite{b:russReport}  The building would be approximately 70~ft $\times$ 80~ft  with a full-span 30~ton bridge crane.  A schematic of the building location and adjacent parking area are found in Figure~\ref{f:g-2:g2Hall}.
The cryogenic needs of the experiment can be met by the Tevatron accelerator cryogenics system with some modifications and additional transfer line work, as the Tevatron is located only about 50~ft away from the AP0 service building.  If necessary, a stand-alone system can be considered as well.

\section{Summary}

As the Tevatron collider era winds down and a new era using a very high intensity proton source continues to develop, an opportune time window exists in which an 8 GeV intensity-frontier physics program can be implemented at Fermilab over the next decade.  The Mu2e Experiment has already achieved Stage-I approval from the laboratory, and will utilize the existing 8 GeV infrastructure with relatively little overall modification to the accelerator complex.  A second initiative -- the New g-2 Experiment -- is seeking approval to use much of this infrastructure as well.  Both experiments, running consecutively, would use spare cycles of the Fermilab Booster to provide an average of 25~kW beam power on target, with no impact on the 400-700~kW neutrino program running concurrently from the Main Injector.  Further updates on these two experiments and the accelerator developments at Fermilab can be found at their web sites, \cite{b:mu2eWeb} and \cite{b:g2Web}.

\bigskip

\begin{acknowledgments}
The author, acting as reporter, would like to acknowledge the many Fermilab staff members and other collaborators of both the Mu2e and g-2 endeavors who have worked, and are working, extensively on these efforts.
\end{acknowledgments}

\bigskip 

\end{document}